\documentclass[a4paper,fleqn]{cas-dc}
\usepackage[noend]{algpseudocode}
 \usepackage{
  booktabs,
  color,
  float,
  svg,
  wrapfig,
  graphics,
  graphicx,
  subcaption,
  algpseudocode,
  tikz,
  xurl,
  breakurl,
  enumitem,
  mfirstuc,
  colortbl
}

\usepackage{soul}
\usepackage{booktabs}
\usepackage[noend]{algpseudocode}
\usepackage{algorithm}
\usepackage{subcaption}
\usepackage{url}
\usepackage{cite}
\usepackage{amsmath,amssymb,amsfonts}
\usepackage{graphicx}
\usepackage{textcomp}

\setlist[itemize]{nosep}

\begin{document}

\let\WriteBookmarks\relax
\def\floatpagepagefraction{1}
\def\textpagefraction{.001}

\shorttitle{Neonpool}    

\shortauthors{HB Haq, ST Ali, A Salman, P McCorry, SF Shahandashti}  

\title [mode = title]{Neonpool: Reimagining Cryptocurrency Transaction Pools for Lightweight Clients and IoT Devices }  



\author[label1]{Hina Binte Haq} 
\author[label1]{Syed Taha Ali}
\author[label2]{Asad Salman}
\author[label3]{Patrick McCorry}
\author[label4]{Siamak F. Shahandashti}

\cormark[1]


\ead{siamak.shahandashti@york.ac.uk}

\credit{
    Hina Binte Haq: Conceptualization, Investigation, Methodology, Visualization, Writing - original draft; 
    Syed Taha Ali: Conceptualization, Supervision, Writing -- review \& editing; 
    Asad Salman: Data curation; 
    Patrick McCorry: Conceptualization; 
    Siamak F. Shahandashti: Validation, Writing -- review \& editing
}

\affiliation[label1]{organization={National University of Sciences and Technology (NUST)}, 
            city={Islamabad}, 
            country={Pakistan}}

\affiliation[label2]{organization={X (formerly Twitter)}, 
            country={USA}}

\affiliation[label3]{organization={Arbitrum}, 
            city={London}, 
            country={United Kingdom}}

\affiliation[label4]{organization={University of York}, 
            city={York}, 
            country={United Kingdom}}

\cortext[1]{Corresponding author}

\nonumnote{A preprint of this paper is available on arXiv.}

\begin{abstract} The increasing adoption of cryptocurrencies has significantly amplified the resource requirements for operating full nodes, creating substantial barriers to entry. Unlike miners, who are financially incentivized through block rewards and transaction fees, full nodes lack direct economic compensation for their critical role in maintaining the network. A key resource burden is the transaction pool, which is particularly memory-intensive as it temporarily stores unconfirmed transactions awaiting verification and propagation across the network. We present \textit{Neonpool}, a novel optimization for transaction pool leveraging bloom filter variants to drastically reduce memory consumption by up to 200x (e.g., 400 MB to 2 MB) while maintaining over 99.99\% transaction processing accuracy. Implemented in C++ and evaluated on unique Bitcoin and Ethereum datasets, \textit{Neonpool} enables efficient operation on lightweight clients, such as smartphones, IoT devices, and systems-on-a-chip, without requiring a hard fork. By lowering the cost of node participation, \textit{Neonpool} enhances decentralization and strengthens the overall security and robustness of cryptocurrency networks. 
\end{abstract}


\maketitle

\section{Introduction}
\label{sec:introduction}
Cryptocurrencies are revolutionizing finance by fostering decentralization, efficient cross-border transactions, and creating new investment opportunities. In recent years, cryptocurrencies have witnessed global impact, gaining users and acceptance by major players like PayPal and Tesla~\cite{bitpay}, and governments actively exploring and piloting central bank digital currencies. Blockchain technology has also driven innovation beyond finance, impacting domains such as healthcare, real estate, freight and supply chains, etc.

Participating in the cryptocurrency ecosystem, however, is a significant undertaking: running cryptocurrency nodes entails growing resource costs (hardware, bandwidth, and electricity consumption). Many novel lightweight cryptocurrency clients have been proposed over the years to address this issue~\cite{chatzigiannis2022sok}. These clients cater to diverse users and typically prioritize certain node functions over others.

For instance, pruned nodes conserve storage by discarding old transactions. Simplified payment verification (SPV) clients, designed for lightweight devices, store block headers and only request transactions of interest from full nodes~\cite{nakamoto2008}. Other proposals include lowering computation costs using lightweight transaction inclusion proofs~\cite{zamyatin2020txchain}, minimizing state size~\cite{chepurnoy2018edrax}, and reducing bandwidth consumption using limited flooding and intermittent reconciliation of transactions~\cite{naumenko2019erlay}. Most light clients cannot function independently and rely on full nodes for proper functioning.


Furthermore, none of the clients proposed thus far cater to the growing local memory (RAM) consumption of cryptocurrency nodes. This includes the transaction pool, which indexes unconfirmed transactions in local memory for inventory purposes and network-wide propagation. The transaction pool uses map data structures to store, manage, and organize transactions, resulting in memory usage significantly greater than the actual transaction data, typically several hundreds of megabytes.  Storing the transaction pool in RAM is two orders of magnitude faster than disk storage~\cite{jiang2019bzip}. 

Increased transaction loads substantially increase pool size, which strains the resources of nodes, and results in dropped transactions, processing delays, spikes in transaction fees, and even exposes the network to sophisticated attacks. Additionally, the transaction pool is also a vector for spam and dust attacks. In October 2015, a Bitcoin spam campaign grew the transaction pool to 1 GB (88k transactions), crashing 10\% of nodes, mostly on Raspberry Pi~\cite{baqer2016stressing}.

In this paper, we propose \textit{Neonpool}, a novel solution that optimizes the transaction pool for resource-constrained platforms. \textit{Neonpool} uses probabilistic data structures to design the transaction pool, which utilizes statistical properties for compact representations of large data sets, offering highly space-efficient solutions that provide answers to membership queries with tightly controlled error rates.

\textit{Neonpool} utilizes two key insights: first, we observe that a majority of light clients levy a burden on the network by piggybacking on existing full-node clients for their proper functioning. Second, we note that the two key functions of the transaction pool, \textit{inventory} and \textit{forwarding}, can be dissociated. This approach is similar to how lightweight clients (e.g., pruned nodes, SPV wallets) are commonly built by prioritizing one function over another \cite{chatzigiannis2022sok}. So we ask the question, is it possible to get the best of both worlds, i.e., design a client that optimizes local memory and contributes to the network's health without imposing a significant burden on existing full nodes?

Specifically, we make the following contributions:
\begin{itemize}
    \item We describe \textit{Neonpool}, an optimized transaction pool construction for cryptocurrencies that explores using standard bloom filters, decaying bloom filters, and bloom filters with key-value storage to replicate the transaction pool's core function of transaction inventory. 
    It reduces the transaction pool's local memory consumption by up to two orders of magnitude (400 MB to 2 MB) while still processing unconfirmed transactions with over 99.99
    
    \item We implement two variants: \textit{Neonpool-BTC} and \textit{Neonpool-ETH}, individually developed in C++ and benchmarked on two novel blockchain network datasets, each containing 10 million unique Bitcoin and Ethereum network transactions.
    
    \item \textit{Neonpool-BTC} and \textit{Neonpool-ETH} are theoretically and empirically evaluated on multiple dimensions, including error rates, memory utilization, computation time, and security on popular IoT devices.
\end{itemize}


To the best of our knowledge, \textit{Neonpool} is the first optimization solution targeted specifically at the transaction pool and local memory. Our results demonstrate a dramatic reduction in memory consumption, up to 200x (400 MB to 2 MB for Bitcoin and Ethereum), with transaction processing accuracy over 99.99\%. This solution enables resource-constrained systems like smartphones, systems-on-a-chip, mobile, and IoT devices to run a high-performing functional transaction pool. It does not require a hard fork and is orthogonal to other light clients. \textit{Neonpool} may be combined with them to aggregate their benefits. \textit{Neonpool} can be extended to other cryptocurrencies. 

\textit{Neonpool} helps reduce the cost of running a full node for users. Running a network node contributes to the health of the network: it helps keep the network decentralized, as each node independently enforces consensus rules and validates and verifies transactions. It also ensures privacy for users, unlike SPV nodes and wallets, which expose transaction history to external servers.

In the subsequent sections, we delve into the requisite background in \S\ref{sec:neonpool_background}, followed by the proposed scheme in \S\ref{sec:neonpool_proposed}. We analyze and discuss empirical results in \S\ref{sec:neonpool_results}. We compare our scheme with prior work in \S\ref{sec:neonpool_prior}. We identify potential future directions and conclude in \S\ref{sec:neonpool_conclusion}.

\section{Background}
\label{sec:neonpool_background}

\subsection{Unconfirmed transaction pool}
\label{sec:neonpool_pool}
The unconfirmed transaction pool (alternately called the transaction pool) serves as a gateway for verifying and temporarily storing unconfirmed or pending transactions in a cryptocurrency node while they await inclusion in a block.

Its primary functions include \textbf{ 1. Transaction verification:} All incoming transactions are checked for adherence to the cryptocurrency's protocol rules such as syntax, valid signatures, availability of funds, etc; \textbf{2. Transaction storage:} A verified transaction is stored temporarily until it is included in a block. \textbf{3. Transaction propagation:} The verified transactions are disseminated through the peer-to-peer network. A verified transaction is only forwarded to the peers the first time it is received by a node to prevent loops in the network. Forwarding a transaction enables other nodes to independently verify and store them in their transaction pool. 

Other functions include estimating transaction fees, prioritizing transactions for inclusion in a block, and bootstrapping the transaction pool of a newly connected node.  


The components of the transaction pool can vary depending on the cryptocurrency but generally include \textbf{Transaction data:} the raw transaction data itself; \textbf{Transaction metadata:} other relevant information associated with each transaction, such as the time the transaction was received, the fee it pays, transaction's priority, to name a few; \textbf{Transaction indexing data:} to efficiently search and retrieve transactions from the transaction pool, implementations utilize indexing structures like maps or priority queues. They enable rapid lookup, insertion, and deletion operations based on transaction IDs or other unique identifiers, to help optimize the process of transaction validation and propagation.

In Bitcoin, the mempool is allocated 300 MB by default~\cite{mempoollimit}. In Ethereum, the default number of pending transactions in the txpool is 4096~\cite{srccodetxpoolgo}. Surplus transactions are evicted. Users can define a custom transaction pool acceptance policy. In a low-memory environment, the transaction pool size can be disabled entirely.

\begin{figure}[htbp]
\centering
    \begin{subfigure}{0.5\textwidth}
    \centering
        \includegraphics[height=3.5cm,width=8cm]{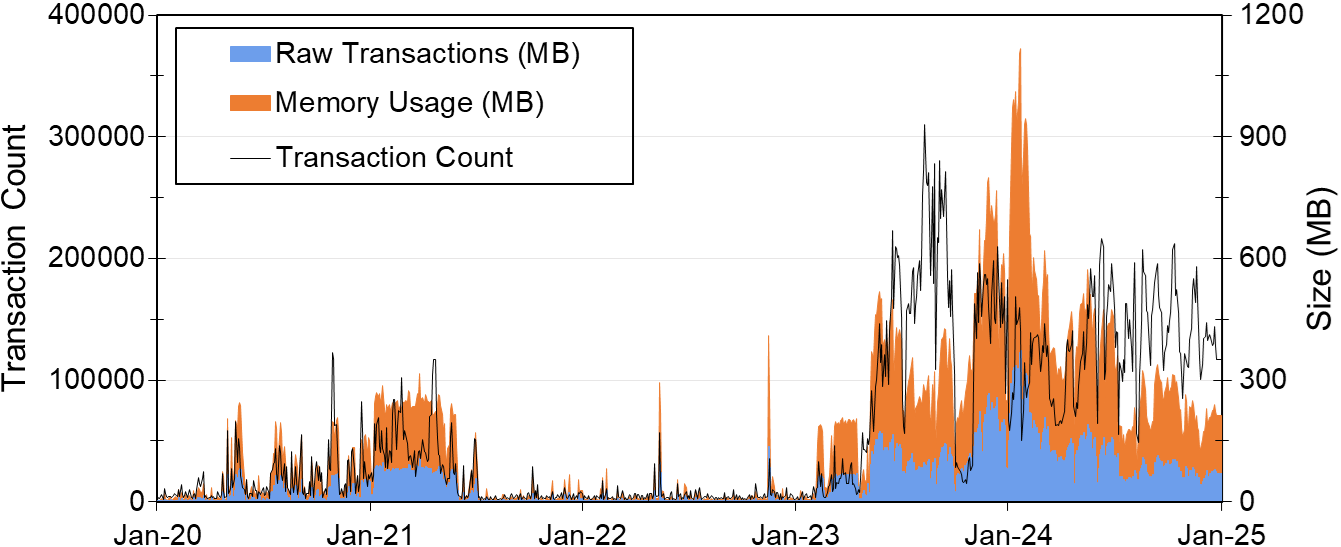}
        \caption{Bitcoin transaction count, raw size~\cite{btccharts} and memory usage}
        \label{fig:neonpool_histmempool}
    \end{subfigure}
    \begin{subfigure}{0.5\textwidth} 
    \centering
    \vspace{2mm}
        \includegraphics[height=3.5cm,width=8cm]{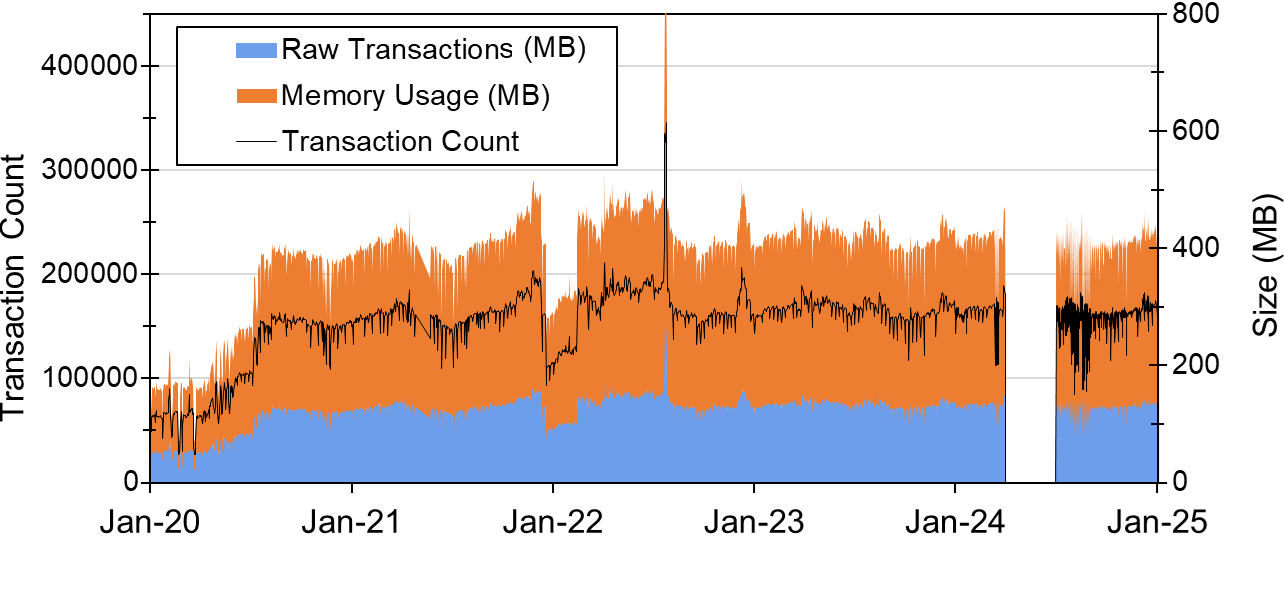}
        \caption{Ethereum transaction count~\cite{etherscan}, raw size and memory usage}
        \label{fig:neonpool_histtxpool}
   \end{subfigure}
    \caption{Transaction pool trends (since 2020)}
     \label{fig:neonpool_hist}
\end{figure}

Fig 1a provides the transaction count and raw transaction size of transactions in the Bitcoin mempool btccharts. Using this data, we calculate the memory usage of the Bitcoin mempool, which is approximately three times larger than the size of the raw transaction data~\cite{300mbproblem}~\cite{Hoenicke2023Apr}. This disproportionate memory usage arises from the inherent limitations of map data structures, which require significant additional memory due to the storage of metadata, indexes, and pointers, resulting in total memory usage that far exceeds the size of the raw data. In early 2024, Bitcoin's transaction pool often ranged from 150-300 MB in raw transaction size, with memory usage exceeding 400 MB.

Etherscan provides comprehensive data on transaction count for the Ethereum txpool~\cite{etherscan}, as shown in Fig 1b. As explained above, overheads are estimated to be three times the raw transaction size for Bitcoin.\footnote{Simple ETH transfers are 30\% of Ethereum transactions (210–250 bytes), while smart contract transactions make up 70\% (500–1,000 bytes), resulting in an average size of ~600 bytes.~\cite{glassnode2024Dec}} This helps estimate memory usage. These metrics are visualized in Fig. 1b. For Ethereum, raw transaction size has frequently surpassed 150 MB, while memory consumption often crosses 400 MB.

\begin{figure}[htbp]
    \centering
    \includegraphics[height=1.9cm, width=8.5cm]{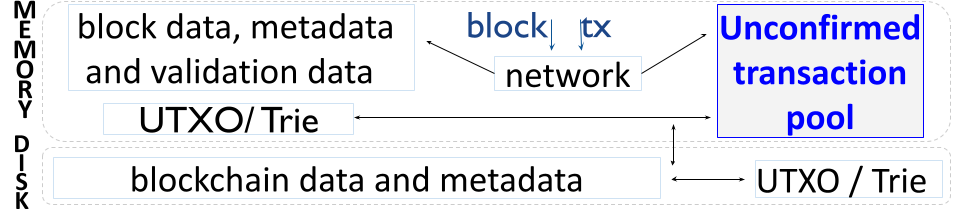}
    \caption{Major node components in disk and memory}
    \label{fig:neonpool_bloomschematic}
\end{figure}

Fig.~\ref{fig:neonpool_bloomschematic} shows the distribution of node components over the hard disk and RAM. Disk storage encompasses raw block data, metadata, and state information like UTXO or Trie. Notably, the UTXO/Trie is partially mirrored in RAM. Additionally, RAM contains essential elements such as the unconfirmed transaction pool, partial state (e.g., UTXO or Trie), block and validation cache, as well as network connections information. 

\begin{figure}[htbp]
    \centering
    \includegraphics[width=8.5cm]{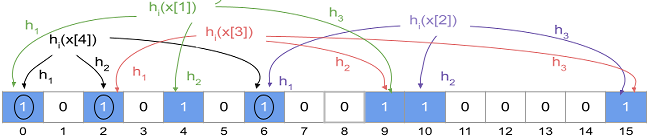}
    \caption{Bloom filter}
    \label{fig:neonpool_bf}
\end{figure}

\subsection{Bloom filter}
\label{sec:neonpool_bloomfilter}
A Bloom filter~\cite{bloom1970space} is a probabilistic data structure used to test for set membership. It is particularly efficient with regards to memory usage but with a caveat: it may provide false positive results, i.e., it can report that an element is present in the set when it is not, but it will never produce a false negative.

Essentially it is a bit array of size \emph{m} and \emph{k} different hash functions. Initially, all bits are set to 0. To insert an element, it is hashed using each of the \emph{k} hash functions, resulting in \emph{k} different bits corresponding to positions in the bit array, which are set to 1. To check for membership, the element is again hashed by the \emph{k} hash functions, and if any of the corresponding positions in the bit array are 0, the element is not in the set. If all the positions are 1, the element is likely in the set. In Fig.~\ref{fig:neonpool_bf} we insert three elements $x[1], x[2]$, $x[3]$ in the filter, which map to indices $\{2,9,15\}$, $\{6,10,15\}$ \& $\{0,4,9\}$ respectively. For $x[4]$ with indexes $\{0,2,6\}$, the bloom filter returns a \emph{false positive} due to earlier insertions setting those bits to 1. The probability of a false positive, alternatively called the false positive rate (FPR), depends on the size of the bit array, the number of hash functions used, and the number of elements inserted. It is given by 
\begin{eqnarray}
\label{eq:fpr}
  p & \approx & \left( 1-e^{-\frac{kn}{m}} \right)^k 
\end{eqnarray}
\textit{FPR} highlights the trade-off between space and accuracy. Filter size $m$ can be provisioned as per set size $n$: 
\begin{eqnarray}
\label{eq:size}
m & \approx & \left( -n \times {\frac{ln (p)}{(ln 2)^2}} \right)
\end{eqnarray}
The optimum value of a number of hash functions, $k$ is 
 \begin{eqnarray}
 \label{eq:hash}
k & \approx & \left ({\frac{m}{n}} \times ln 2 \right)
\end{eqnarray}

\textbf{Decaying Bloom filters}~\cite{bianchi2011demand} randomly decay or "age out" bits over time. This behaviour is helpful when item relevance decreases over time, and it’s important to maintain an accurate representation of recent data while letting old data expire. Upon inserting an element into the Decaying Bloom filter, bits are reduced by a constant decay factor. A Decaying Bloom filter can be achieved by modifying a standard Bloom filter, whereupon every insertion $d$ random indices are decremented, mimicking expiry. For instance, the decay factor, $d=32$, results in 32 random indices of the filter being decremented upon every insertion. A fraction of these randomly selected indices may be already zero.  

Bloom filters are commonly used in applications, such as caching and network routing, where probabilistic results and memory efficiency are acceptable trade-offs. 

\begin{figure*}[t]
\begin{minipage}[t]{0.48\linewidth}
\centering
\begin{algorithm}[H]
\caption{\textit{transaction pool}}
\begin{algorithmic}[1]
\Function{receiveTransaction}{tx}
    \If{\Call{transactionpoolLookup}{tx}}
        \State \Call{dropTransaction}{tx}
    \Else
        \If{\Call{syntaxAndSemanticsCheck}{tx}}
            \If{\Call{doublespendCheck}{tx}}
                \State \Call{addTransaction}{tx}
                \State \Call{relayTransaction}{tx}
            \Else
                \State \Call{dropTransaction}{tx}
            \EndIf
        \Else
            \State \Call{dropTransaction}{tx}
        \EndIf
    \EndIf
\EndFunction
\end{algorithmic}
\label{alg:txpool}
\end{algorithm}
\end{minipage}\hfill
\begin{minipage}[t]{0.48\linewidth}
\centering
\begin{algorithm}[H]
\caption{\textit{Neonpool}}
\begin{algorithmic}[1]
\Function{receiveTransaction}{tx}
    \If{\Call{bloomtxFilter}{tx}}
        \State \Call{dropTransaction}{tx}
    \Else
        \If{\Call{syntaxAndSemanticsCheck}{tx}}
          \If{\Call{dstxFilter}{tx}}
                \State \Call{addTransaction}{tx}
                \State \Call{relayTransaction}{tx}
            \Else
                \State \Call{dropTransaction}{tx}
            \EndIf
        \Else
            \State \Call{dropTransaction}{tx}
        \EndIf
    \EndIf
\EndFunction
\end{algorithmic}
\label{alg:neonpool}
\end{algorithm}
\end{minipage}
\end{figure*}

\textbf{Bloom filters in cryptocurrencies:}Transaction Bloom filtering, introduced in BIP 37, enabled lightweight SPV wallets to efficiently retrieve transactions relevant to their addresses by sending a Bloom filter (based on its addresses, public keys, or other identifiers) to a full node, reducing bandwidth usage. However, BIP 37 was deprecated in Bitcoin Core 0.21.0 (2021) due to privacy concerns, as it exposed SPV wallet addresses to the full node. Graphene uses bloom filters to reduce network bandwidth in block reconciliation in Bitcoin~\cite{ozisik2019graphene}. In Ethereum, Bloom filters are used within block headers to summarize the logs generated by transactions in a block. This enables quick event log lookups without requiring full transaction execution. Bloom filters support block declaration fairness by enabling nodes to quickly verify transaction inclusion when competing blocks are declared, ensuring valid transactions are not overlooked. For a comprehensive summary of the applications of bloom filters in blockchain systems, we refer the reader to~\cite{rottenstreich2021sketches}.

A Merkle tree is a binary tree structure used in cryptography to efficiently and securely verify the integrity of large datasets. They are invaluable for ensuring data integrity and inclusion in static, immutable datasets. Bitcoin and Ethereum leverage Merkle trees to organize transaction data within blocks. Once a block is mined and added to the chain, its data becomes immutable, making Merkle trees a suitable choice for this static context. However, their high update costs, verification overhead, and reliance on additional data (e.g., root and sibling hashes) make them unsuitable for the dynamic, high-churn environment of transaction pools.

In contrast, Bloom filters are particularly suited for optimizing the transaction pool because they enable efficient insertion and verification without requiring additional data, such as sibling or root hashes. Unlike Merkle trees, which scale logarithmically with the size of the dataset, the memory usage of a Bloom filter is independent of the dataset size. Instead, it is determined by its configuration, such as the size of the bit array and the number of hash functions. This makes Bloom filters especially beneficial for resource-constrained environments, including lightweight clients and IoT devices.

\section{Proposed scheme}
\label{sec:neonpool_proposed}

In this section, we provide a detailed explanation of our proposed scheme. By leveraging probabilistic data structures, \textit{Neonpool} maintains the transaction pool's core functionality while significantly improving resource efficiency. It does not necessitate the storage of complete transactions. Instead, it only stores transaction fingerprints, effectively disassociating the processes of transaction forwarding and inventory management.

\textit{Neonpool} comprises two primary components. The first, \texttt{bloomtxFilter}, utilizes the transaction ID or hash, \texttt{txHash}, to map valid ingress transactions. The second component, \texttt{dstxFilter}, ensures that duplicate or potential double-spend transactions are identified and discarded. The exact mechanism to identify potential double-spending varies for UTXO-based Bitcoin and account-based Ethereum. 

We present two variations of our proposed scheme, namely \textit{Neonpool-BTC} and \textit{Neonpool-ETH}. The term \textit{Neonpool} is used if a certain aspect of the scheme applies to both Bitcoin and Ethereum. We use the term \textit{mempool} to refer to the Bitcoin transaction pool and \textit{txpool} to refer to the Ethereum transaction pool, while \textit{transaction pool} is a generic term that applies to both \textit{Bitcoin mempool} and \textit{Ethereum txpool}.

\subsection{Ingress}
Here, we describe the ingress process for the transaction pool and \textit{Neonpool}, as depicted in Algo.~\ref{alg:txpool} and~\ref{alg:neonpool}, respectively.

\begin{figure}[ht]
\centering
    \begin{subfigure}{0.5\textwidth}
    \centering
         \includegraphics[width=8cm]{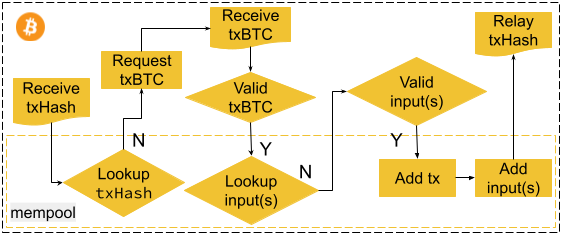}
    \caption{Mempool Ingress}
    \label{fig:neonpool_neonpoolbtc}
    \end{subfigure}
    \begin{subfigure}{0.5\textwidth} 
    \centering
    \vspace{2mm}
      \includegraphics[width=8cm]{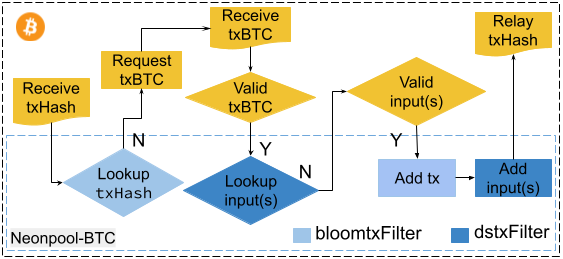}
    \caption{\textit{Neonpool-BTC} Ingress}
    \label{fig:neonpool_neonpoolbtc1}
   \end{subfigure}
    \caption{Ingress: mempool vs \textit{Neonpool-BTC} (only (Y)es or (N)o is shown, other implies transaction is dropped.)}
     \label{fig:neonbtc}
\end{figure}

\subsubsection{\textbf{Neonpool-BTC}} 
As shown in Fig.~\ref{fig:neonpool_neonpoolbtc} and~\ref{fig:neonpool_neonpoolbtc1}, in both Bitcoin Core and \textit{Neonpool-BTC}, the process begins with the arrival of a transaction announcement through an \textit{inv} message. In Bitcoin, \texttt{txHash} is used to query the \textit{mempool} to determine if the transaction already exists in the \textit{mempool}. In \textit{Neonpool-BTC}, the \texttt{txHash} is used to query the \texttt{bloomtxFilter}.

Nodes may receive a transaction announcement multiple times but only accept it the first time they receive it. This functionality is essential in cryptocurrency networks to minimize traffic overhead and prevent infinite loops by ensuring that transactions are broadcast only upon first receipt. In both Bitcoin and \textit{Neonpool-BTC}, if a transaction with the same \texttt{txHash} has already been received, it is discarded. If the transaction is determined to be new, the complete transaction is requested via a \textit{transaction} message. When received, the complete transaction \textit{txBTC} undergoes syntax, validity (valid transaction signatures, availability of sufficient funds, etc.), and semantics checks in both Bitcoin and \textit{Neonpool-BTC}, followed by checks to detect double-spends.

In both Bitcoin and \textit{Neonpool-BTC}, each of the \textit{inputs}, comprising the \texttt{inputtxHash} and \texttt{index} are scanned for double-spends. Transaction inputs are validated using the UTXO set. Transactions with invalid or spent inputs are discarded. It is also checked that none of the inputs exists in the \textit{ mempool} in Bitcoin. For \textit{Neonpool-BTC} the \texttt{dstxFilter} is queried with the tuple \texttt{<inputtxHash, index>} to ascertain that the input does not already exist in the filter. If any of the inputs already exist in the \textit{mempool} or \texttt{dstxFilter}, the transaction is dropped, as it constitutes a potential double spend. If two transactions with the same inputs are in circulation, the first seen by a node is regarded as safe, while the second is dropped. If any transaction input also (referred to as \textit{parent} or \textit{ancestor}) is missing, the transaction is added to the orphan pool. It will reside in the orphan pool until its ancestor is received, after which it will be reprocessed. If the \textit{txBTC} passes the verification process, it is added to the \textit{mempool} in Bitcoin and \texttt{bloomtxFilter} in \textit{Neonpool-BTC}. Finally, the transaction hash, \texttt{txHash}, is relayed to the connected peers. 

\begin{figure}[ht]
\centering
    \begin{subfigure}{0.5\textwidth}
    \centering
        \includegraphics[width=8cm]{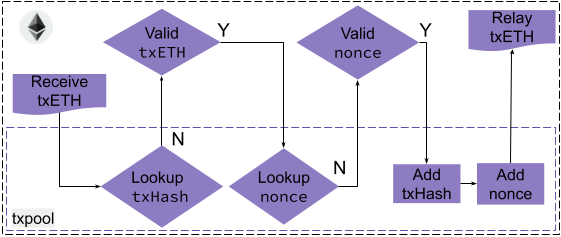}
    \caption{Txpool Ingress}
    \label{fig:neonpool_neonpooleth}
    \end{subfigure}
    \begin{subfigure}{0.5\textwidth} 
    \centering
    \vspace{2mm}
      \includegraphics[width=8cm]{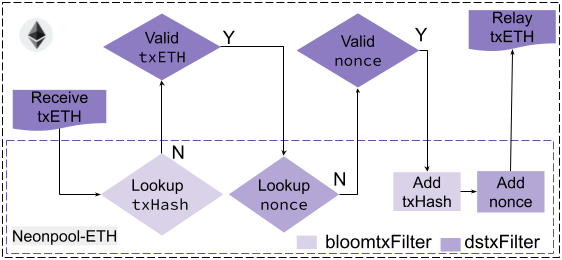}
    \caption{\textit{Neonpool-ETH} Ingress}
    \label{fig:neonpool_neonpooleth1}
   \end{subfigure}
    \caption{Ingress: txpool vs \textit{Neonpool-ETH} (only (Y)es or (N)o is shown, other implies transaction is dropped.)}
     \label{fig:neoneth}
\end{figure}

\subsubsection{\textbf{Neonpool-ETH}}
As shown in Fig.~\ref{fig:neonpool_neonpooleth} and ~\ref{fig:neonpool_neonpooleth1}, in both Ethereum and \textit{Neonpool-ETH} a complete transaction \textit{txETH} arrives via a \textit{transaction} message. In Ethereum, the transaction hash (or ID) \texttt{txHash} is used to query the \textit{txpool} to determine if the transaction already exists in the \textit{txpool}. In \textit{Neonpool-ETH}, the \texttt{txHash} is used to query the \texttt{bloomtxFilter}.

Nodes may receive a transaction announcement multiple times but only accept it the first time they receive it. In both Ethereum and \textit{Neonpool-ETH}, if a transaction with the same \texttt{txHash} has already been received, it is discarded. If the transaction is determined to be new, it undergoes syntax, validity (valid transaction signatures, funds availability, etc.), and semantics checks in both Ethereum and \textit{Neonpool-ETH}. 

In both Ethereum and \textit{Neonpool-ETH}, the transaction is then checked for potential double-spends by validating the transaction's \texttt{address, nonce} and \texttt{amount} against the \textit{Trie}. Transactions with an invalid or out-of-order nonce or insufficient funds are rejected. 

Then, in Ethereum, the \textit{txpool} is queried to check if a transaction with the same \textit{address} and \textit{nonce} already exists in it. If two transactions from the same \textit{address} and with the same \textit{nonce} are in circulation, the first seen by a node is regarded as safe, while the second is dropped. The first seen may differ for nodes on the network. In \textit{Neonpool-ETH}, the \texttt{<address, nonce>} tuple of \textit{txETH} is scanned in \texttt{dstxFilter} to detect any potential double-spends and drop such transactions.

In Ethereum, if \textit{txETH} passes the verification process, it is added to the \textit{txpool}. In \textit{Neonpool-ETH}, the \texttt{txHash} is used to add the transaction to the \texttt{bloomtxFilter}, while the tuple \texttt{<address, nonce>}, representing the sender's address, and the nonce value of the transaction is added to the \texttt{dstxFilter}. Finally, the complete transaction is relayed to a random fraction of connected peers. 

\subsection{Egress}
In Bitcoin core and Ethereum, transactions are removed from the unconfirmed transaction pool for various reasons, such as inclusion in a block, limited transaction pool capacity, transaction expiry, fee priority, running out of gas (in Ethereum), replacement by a newer version that offers a higher fee, invalid or conflicting transaction, or chain reorganization event at the node.

In Bitcoin, transactions are automatically removed from the mempool after a default expiration period of 14 days, although this limit can be configured by the node operator~\cite{txmempoolsourcecode}. In Ethereum, the transaction pool's memory usage is configurable, and transactions expire when the number of pending transactions exceeds the default limit of 4,096 or when they remain unprocessed for more than 3 hours~\cite{srccodetxpoolgo}.

\textit{Neonpool} mimics these expiry mechanisms using Bloom filters, offering nodes configurable parameters for transaction expiry through periodic clearing and decaying Bloom Filters. This is necessary because we do not perform any deletions on the bloom filter, and the load in the bloom filter is bound to exceed the number of transactions it was initially dimensioned for. We consider two approaches: Firstly, the accumulated transactions can be periodically removed by \textbf{clearing the filter}. This period may be a fixed hourly interval or based on the number of transactions processed. Secondly, Decaying Bloom filters may be employed to \textbf{randomly decay or age out transactions}, to maintain an accurate representation of recent data while letting old data expire. 

Our work is a pioneering investigation into the feasibility and suitability of probabilistic data structures for transaction pool construction. Based on the core function of the transaction pool, we require a data structure with the following properties: it can answer membership queries, let old transactions expire to make way for more recent ones and ensure double spending protection. Key-value support is also required. While there are more than a dozen variants of bloom filters available in the literature, we start by using the most basic ones that meet our requirements and are widely understood. As a proof-of-concept, we evaluate these in \S~\ref{sec:neonpool_results} and gain greater insight into the theoretical limitations of each data structure. Further optimizations will be explored in an extension study.



\subsection{Scaling \textit{Neonpool}} We present a strategy to enable \textit{Neonpool} to handle increasing transaction loads effectively: To mitigate the risk of false positives during heavy network congestion, the system initializes a counter to track the total number of transactions inserted into the Bloom filter. When the transaction count exceeds the capacity of 200,000, additional auxiliary Bloom filters of the same size and capacity are dynamically instantiated, as per demand and expired in order of age~\cite{guo2009dynamic}\cite{beyer2011system}. This prevents overloading the filter, thereby maintaining acceptable false positive rates.

The proposed approach of recursively generating Bloom filters may result in increased computation time and memory overhead on lightweight devices. In cryptocurrencies like Bitcoin and Ethereum, there is an established practice of limiting transaction pool sizes and managing overflow by rejecting or expiring excess transactions. For instance, Bitcoin’s default mempool size is capped at 300 MB, while Ethereum employs a default limit of 4096 transactions in the transaction pool, with surplus transactions being evicted~\cite{mempoollimit}~\cite{srccodetxpoolgo}. Furthermore, users can customize transaction pool policies or disable transaction pools entirely to accommodate low-memory or low-computation environments.  

Similarly, Neonpool allows nodes to configure and adjust memory and computational thresholds based on their specific capacity. If the memory or computational threshold is reached, older Bloom filters automatically expire as necessary. For example, if a node can efficiently manage only two concurrent filters without exceeding its computational resources, it can set this limit to ensure optimal performance.

The objective is to empower full-node users by optimizing resource utilization, providing enhanced flexibility, and granting greater control over resource allocation. This approach incentivizes altruistic participation of the full nodes, enabling them to actively participate in transaction verification and forwarding, thus contributing to the decentralization, security, and robustness of the ecosystem.

The scaling function of Neonpool is governed by Eq.~\ref{eq:size} and is determined by the configuration of the Bloom filter, specifically the size of the bit array and the number of hash functions:
\textbf{False-Positive Rate (p):} Eq. 2 shows that the bit array size (m) grows logarithmically with p. For example, reducing the false-positive probability from $10^{-3}$ to $10^{-6}$ increases the bit array size modestly. This logarithmic relationship allows for efficient use of resources while lowering false positives.
\textbf{Number of Transactions (n):} Assuming the false-positive rate (p) is kept constant, the bit array size grows linearly with the number of transactions. As the number of transactions (n) increases, the size of the Bloom filter increases proportionally. If the number of transactions doubles, the Bloom filter size doubles, making it scalable as transaction pools grow.

Thus, Neonpool scales logarithmically with the desired false-positive probability and linearly with the number of transactions it tracks, ensuring efficient transaction pool management even in resource-constrained environments.

\section{Experiments, results and discussion}
\label{sec:neonpool_results}
This section comprehensively evaluates \textit{Neonpool-BTC} and \textit{Neonpool-ETH} in comparison to \textit{Bitcoin mempool} and \textit{Ethereum txpool} respectively, on multiple dimensions including error rates, memory utilization, computation time, and security, on popular IoT devices.

\subsection{Data set, implementation, and methodology}
\label{sec:neonpool_dataset}
We record $ingress$ and $egress$ transactions in the transaction pool in JSON format for Bitcoin and CSV for Ethereum (for raw transaction structure in Bitcoin and Ethereum, see~\cite{rawtxbtc}~\cite{rawtxeth}) to allow us to reconstruct the transaction pool state at the client and replay network activity for simulation purposes. Our data set also includes all transactions (for Bitcoin inventory and Ethereum transaction message structure see~\cite{invmsgbtc}~\cite{txmsgeth}) received over the network stored in CSV format. For Bitcoin, we run an instrumented version of Bitcoin Core modifying \texttt{txmempool.cpp} to capture 10 million unique transactions (around 30 million transaction announcements over $\sim$30 days). Similarly, for Ethereum, we run an instrumented version of Geth, modifying \texttt{txpool.go}, to capture 10 million unique transactions (around 13 million transactions over $\sim$10 days).

We develop simulations for \textit{Bitcoin mempool} and \textit{Ethereum txpool} using map data structures, with high-level pseudocode described in Algo~\ref{alg:neonpool}, and for \textit{Neonpool-BTC} and \textit{Neonpool-ETH} as shown in Fig.~\ref{fig:neonpool_neonpoolbtc} and~\ref{fig:neonpool_neonpooleth}. We replay transactions in each dataset to reconstruct the \textit{Bitcoin mempool} and \textit{Ethereum txpool} over 30 and 10 days, respectively. The simulated \textit{Bitcoin mempool} and \textit{Ethereum txpool} serve as the ground truth, running in parallel with \textit{Neonpool-BTC} and \textit{Neonpool-ETH} to evaluate our scheme's performance.

Neonpool-BTC was implemented using C++, as Bitcoin Core, the most widely used Bitcoin client, is written in C++. Furthermore, C++ offers a robust standard library optimized for high performance, efficient memory management, and handling data-intensive operations with minimal overhead. For Bloom filters and variants, we use the comprehensive Berkeley libbf library~\cite{libbf} written in C++ and incorporate the $H3_{exp}$ hash functions.  

For \textit{Neonpool-ETH}, we reuse the code components developed for \textit{Neonpool-BTC}. The language-agnostic nature of our approach ensures that \textit{Neonpool}'s advantages are preserved even when ported to Go for integration with Ethereum. Our dataset and code are publicly accessible.~\cite{neonpool2024}

We perform independent queries on \textit{Neonpool} and the transaction pool at each $ingress$ transaction in our data set, and the responses are recorded. The responses may diverge from the ground truth owing to the probabilistic nature of bloom filters. We discuss how these false positives and negatives affect our scheme and offer a quantitative analysis. 

\subsection{Evaluation metrics}
The responses obtained from the \texttt{bloomtxFilter} can be categorized as \textbf{True Positive (TP)}: a positive instance correctly classified as positive; \textbf{True Negative (TN)}: a negative instance correctly classified as negative; \textbf{False Positive (FP)}: a negative instance incorrectly classified as positive.

\begin{table*}[t]
\setlength{\tabcolsep}{2pt}
\centering
\begin{tabular}{lrrrrrr}
\toprule
\textbf{Expiry} & \multicolumn{3}{c}{\textbf{Rejected Transactions}} & \multicolumn{3}{c}{\textbf{Redundant Transactions}} \\
\textbf{Hours(h)/} & \textbf{500 kB} & \textbf{1 MB} & \textbf{2 MB} & \textbf{500 kB} & \textbf{1 MB} & \textbf{2 MB} \\
\textbf{Decay(d)} &\textbf{FPR/Num} & \textbf{FPR/Num} & \textbf{FPR/Num} & \textbf{FNR/Num} & \textbf{FNR/Num} & \textbf{FNR/Num} \\
\midrule
\textbf{None} & 8.06E01/26897923 & 7.43E01/25191220 & 6.82E01/23508643 & 0/0 & 0/0 & 0/0 \\
\textbf{h=48} & 5.20E-02/1650093 & 2.08E-02/658984 & 3.77E-03/119770 & 1.08E-03/34309 & 1.17E-03/37060 & 1.21E-03/38510 \\
\textbf{h=24} & 9.03E-03/286612 & 1.94E-03/61577 & 6.22E-04/19735 & 1.77E-03/56112 & 1.79E-03/56905 & 1.80E-03/57057 \\
\textbf{h=12} & 2.83E-03/89922 & 9.28E-04/29479 & 6.05E-04/19200 & 1.99E-03/63224 & 2.00E-03/63473 & 2.00E-03/63549 \\
\textbf{h=6} & 1.51E-03/48011 & 7.32E-04/23247 & 5.82E-04/18476 & 2.27E-03/72025 & 2.27E-03/72180 & 2.27E-03/72194 \\
\textbf{h=3} & 1.03E-03/32575 & 6.51E-04/20680 & 5.80E-04/18415 & 3.06E-03/97205 & 3.06E-03/97320 & 3.07E-03/97333 \\
\textbf{400k tx} & 9.80E-03/408250 & 1.70E-03/73000 & 6.00E-03/9824 & 1.49E-03/46785 & 1.50E-03/46813 & 1.51E-03/46872 \\
\textbf{d=16} & 1.73E-03/72561 & 7.27E-04/30427 & 4.87E-04/20398 & 4.64E-03/148179 & 4.66E-03/148905 & 4.67E-03/149097 \\
\textbf{d=32} & 3.85E-04/16112 & 7.44E-05/3115 & 1.90E-05/794 & 5.04E-03/164708 & 5.04E-03/164608 & 5.06E-03/165318 \\
\textbf{d=64} & 1.45E-04/6055 & 2.64E-05/1103 & 4.44E-06/186 & 5.48E-03/182864 & 5.47E-03/182592 & 5.48E-03/182790 \\
\textbf{d=128} & 1.19E-05/498 & 8.53E-06/357 & 1.98E-06/83 & 6.06E-03/207234 & 6.05E-03/206961 & 6.03E-03/206087 \\
\textbf{d=256} & 1.19E-05/497 & 4.35E-06/182 & 8.60E-07/36 & 7.03E-03/247948 & 7.04E-01/248219 & 7.04E-03/248218 \\
\bottomrule
\end{tabular}
\caption{\textit{Neonpool-BTC} performance for n=400k}
\label{tab:neonpoolbtc}
\end{table*}

In the context of an $ingress$ event, the following implications hold $\mathbf{TP_{ingress}}$: the transaction already exists in the pool and will be discarded correctly. $\mathbf{TN_{ingress}}$: the transaction is new and will be added to the pool as intended. $\mathbf{FP_{ingress}}$: the transaction is new and should be added to the pool, but it will be erroneously rejected. Inherently, bloom filters do not have false negatives. However, because we periodically expire older transactions, we may receive a false negative response in our scenario. $\mathbf{FN_{ingress}}$: The transaction has already been added and expired, but it will be erroneously added again. 

Hence, the criteria used to assess the performance are: \\\textbf{False Positive Rate (FPR)} measures the proportion of transactions rejected erroneously, calculated as $\frac{FP_{ingress}}{Queries_{ingress}}$; \\\textbf{False Negative Rate (FNR)} measures the proportion of transactions reprocessed, calculated as $\frac{FN_{ingress}}{Queries_{ingress}}$.

\subsection{Error rates and memory utilization}
\subsubsection{\textit{Neonpool-BTC}}
The highest transaction volumes observed in the \textit{Bitcoin mempool} to date are around 200k. We dimension \texttt{bloomtxFilter} to handle double that, i.e., 400k transactions, because \textit{Neonpool} delays removing transactions, as discussed below. Using Eq.~\ref{eq:size} and Eq.~\ref{eq:hash}, we dimension filters of size 500\,KB, 1\,MB, and 2\,MB with 4M, 8M, and 16M cells, having 7, 14, and 28 hash functions, and theoretical FPR of $8.2E-03$, $6.7E-05$, and $5.0E-09$ respectively.

We replay transaction events in the Bitcoin dataset. When \textit{Neonpool-BTC} runs without a transaction expiry mechanism, transactions accumulate and quickly surpass the filter design capacity. Due to filter overloading, we get poor results. As shown in Tab.~\ref{tab:neonpooleth}, the 500\,KB, 1\,MB, and 2\,MB filters report a false positive rate (FPR) of $8.06E01$, $7.43E01$, and $6.82E01$, erroneously \textit{rejecting} 80.6\%, 74.3\%, and 68.2\% of transactions respectively. However, as no transactions expire, the false negative rate (FNR) is zero.

We introduce expiry mechanisms to prevent the filters from overloading. We follow two approaches: 1. reset the bloom filter at fixed hourly intervals or once the count of transactions surpasses the number of transactions the filter was originally dimensioned for i.e. 400k in our case; 2. employ a decaying bloom filter that decrements a certain number of indices at random upon every insertion, hence mimicking expiry.

\begin{figure}[b]
    \centering
        \centering
        \includegraphics[height=4cm, width=8.5cm]{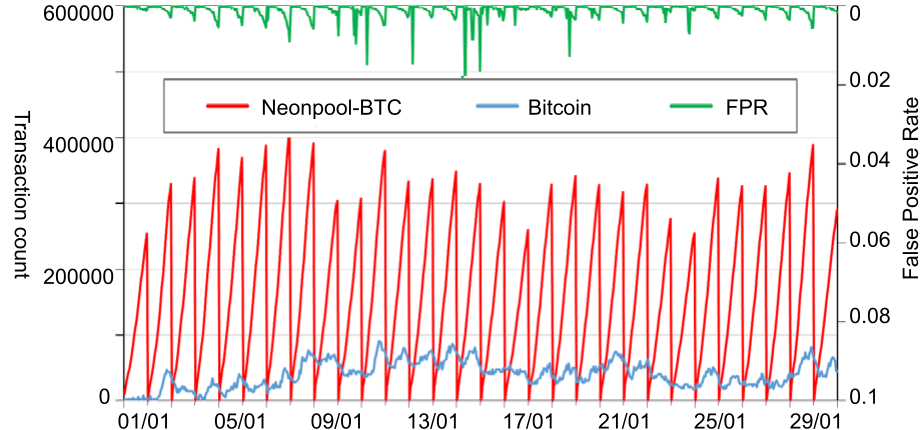}  
        \caption{\textit{Neonpool-BTC} 1\,MB, 24 hours expiry}
        \label{fig:neonpool_btc1}
    \end{figure}
    
Fig.~\ref{fig:neonpool_btc1} depicts in real-time the number of transactions stored in \textit{Neonpool-BTC} with 1\,MB filters and 24-hour expiry alongside the FPR for 30 days (10 million transactions). We also plot the corresponding number of transactions in the \textit{Bitcoin mempool}, the ground truth in our evaluation. The number of transactions closely tracks the pattern in the \textit{Bitcoin mempool}, with an increasing offset, as the filter retains egress transactions until the expiry interval lapses.

For instance, when the filter is cleared every 24 hours the average FPR, at $9.03-03$ is highest for the 500\,kB filter, reducing to $1.94E-03$ and $6.22E-04$ as the filter size increases to 1\,MB and further to 2\,MB. For the 500\,kB, 1\,MB, and 2\,MB filters, this translates to 286612 or $0.90\%$, 61577 or $0.19\%$ and 19735 or $0.06\%$ of transactions being erroneously \emph{rejected} due to false positives. For each filter, there are around $0.18\%$ \emph{redundant} transactions due to false negatives. Tab.~\ref{tab:neonpoolbtc} shows that as the expiry interval is reduced, the FPR improves, while the FNR deteriorates. 

Tab.~\ref{tab:neonpoolbtc} shows that when the filter is cleared every 400k transactions, the average false positive rate at $9.80E-03$ is highest for the 500\,kB filter and reducing to $1.7E-03$ and $6.00E-03$ as the filter size increases to 1\,MB and further to 2\,MB. For the 500\,kB, 1\,MB, and 2\,MB filters. We observe $0.98\%$, $0.17\%$, and $0.06\%$ of transactions being erroneously \emph{rejected} due to false positives. For each filter, there are around $0.15\%$ \emph{redundant} transactions due to false negatives. 

Empirical false positive rates are significantly higher than the theoretical value, sometimes even more than an order of magnitude. We theorize the causes: first, multiple works have reported that false positive rates in real deployments are higher than theoretically computed~\cite{mullin1983second}~\cite{gremillion1982designing}. Researchers contend that this is because theoretical calculations assume that "each hash transformation is perfect"~\cite{mullin1983second} and that transactions "are independent and uniformly distributed over all records" whereas real activity tends to be "clumped"~\cite{gremillion1982designing}. In this context, Bose et al. prove that Eq.~\ref{eq:fpr} gives us a lower bound on the false positive rate~\cite{bose2008false}.

Hence, these deviations reflect practical realities rather than flaws in the theoretical framework. To address this, an effective strategy is to over-dimension the Bloom filter. This compensates for empirical deviations by ensuring false positive rates remain within acceptable bounds for specific applications. Such an approach aligns with established deployment practices, where configurations are optimized based on observed performance metrics rather than idealized theoretical models.

If we use a decaying bloom filter, we achieve vast improvements in terms of false positive rates. The average FPR at $1.19E-05$ is highest for the 500\,kB filter and reduces to $8.53E-06$ and $1.98E-06$ as the filter size increases to 1\,MB and further to 2\,MB. For a decay factor of 128, the 500\,kB, 1\,MB, and 2\,MB filters observe 498 or $0.0012\%$, 357 or $0.0009\%$ and 83 or $0.0002\%$ of transactions being erroneously \emph{rejected} due to false positives and there are around $0.61\%$ \emph{redundant} transactions due to false negatives. Tab.~\ref{tab:neonpoolbtc} shows that by increasing the decay factor, the FPR and, hence, the number of erroneously rejected transactions reduce. This is because the decay average meets the insertion average, and the filter reaches a stable state. However, on the flip side, the false negative rate increases.

The \texttt{dstxFilter} which prevents double spends, can have implications denoted as $\mathbf{TP_{input}}$, $\mathbf{TN_{input}}$, $\mathbf{FP_{input}}$, and $\mathbf{FN_{input}}$. Similar to \texttt{bloomtxFilter}, a $\mathbf{TP_{input}}$ transaction should be discarded, while a $\mathbf{TN_{input}}$ transaction should be accepted. The error $\mathbf{FP_{input}}$ will lead to a genuine transaction being discarded, while $\mathbf{FN_{input}}$ will lead to accepting a transaction the \texttt{<inputtxHash, index>} of which has already been processed. However, circulating such transactions does not imply a double-spend, as \textit{Neonpool-BTC} and other network nodes maintain the UTXO and screen transactions in incoming blocks to prevent double-spending.

\begin{figure}[htbp]
    \centering
    \includegraphics[width=8.5cm, height=4cm]{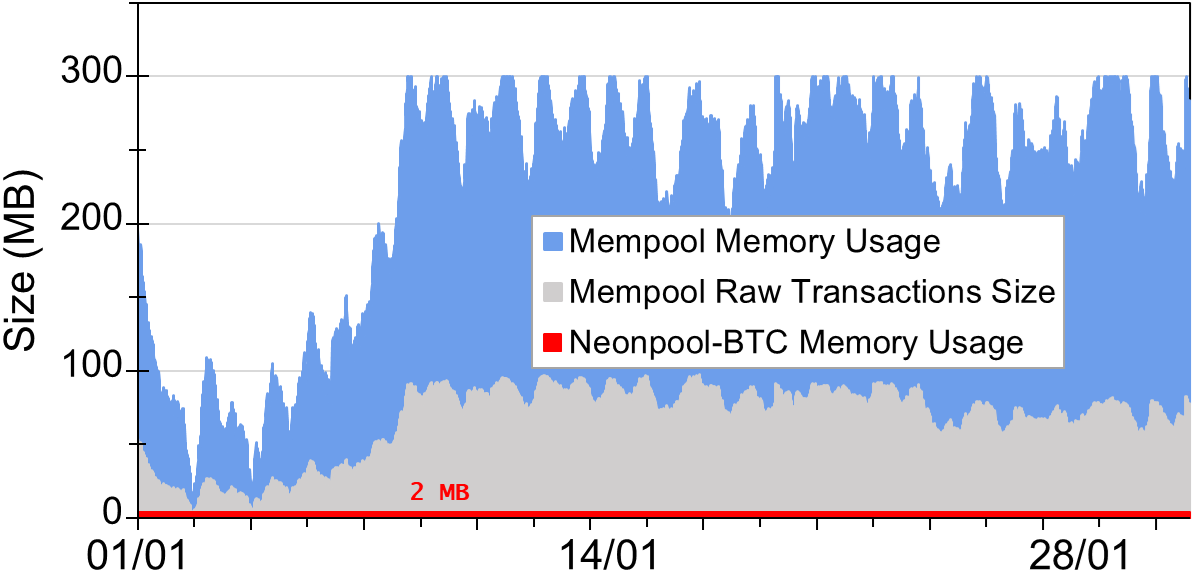}
    \caption{Memory usage of mempool and \textit{Neonpool-BTC}}
    \label{fig:neonpool_neonpoolbtcmemusage}
\end{figure}

In the dataset, incoming transactions average 40,000 inputs hourly, peaking at 191,947 inputs. Thus it is safe that \texttt{dstxFilter} will have the same dimensions as \texttt{bloomtxFilter} and consequently similar FPR. Thus, for a 1 MB \texttt{bloomtxFilter}, rejecting around 0.0009\% of valid transactions, the corresponding \texttt{dstxFilter} will also reject around 0.0009\% of valid transactions. \textbf{Neonpool-BTC achieves 99.99\% fidelity, handling 300 MB of transactions in just 2 MB, as shown in Fig.~\ref{fig:neonpool_neonpoolbtcmemusage}.}

\subsubsection{\textit{Neonpool-ETH}}
The highest transaction volumes observed in the \textit{Ethereum txpool} to date is around 350k. We dimension \texttt{bloomtxFilter} to handle its double i.e. 700k transactions, because \textit{Neonpool} delays the removal of transactions, as discussed below. Using Eq.~\ref{eq:size} and Eq.~\ref{eq:hash} we dimension three filters of size 500\,KB, 1\,MB and 2\,MB, with 4M, 8M and 16M cells, having 4,8 and 16 hashes, and theoretical FPR of $6.4E-02$, $4.1E-03$, $1.7E-05$ respectively.

\begin{table*}[htbp]
\centering
\setlength{\tabcolsep}{2pt}
\begin{tabular}{rrrrrrrrrrrrr}
\toprule
\textbf{Expiry} & \multicolumn{3}{c}{\textbf{Rejected Transactions}} & \multicolumn{3}{c}{\textbf{Redundant Transactions}} \\
 \textbf{event} & \textbf{500 kB} & \textbf{1 MB} & \textbf{2 MB} & \textbf{500 kB} & \textbf{1 MB} & \textbf{2 MB} \\
\textbf{} &\textbf{FPR/Num} & \textbf{FPR/Num} & \textbf{FPR/Num} & \textbf{FNR/Num} & \textbf{FNR/Num} & \textbf{FNR/Num} \\
\midrule
\textbf{None} & 7.30E-01/7932070 & 6.70E-01/7282537 & 6.07E-01/6596909 & 0.00E+00/0 & 0.00E+00/0 & 0.00E+00/0 \\
\textbf{h=48} & 2.33E-01/2529181 & 1.05E-01/1142566 & 2.95E-02/320771 & 1.05E-02/113631 & 1.26E-02/136556 & 8.53E-03/92649 \\
\textbf{h=24} & 5.62E-02/610328 & 8.14E-03/88437 & 5.69E-04/6181 & 1.19E-02/129414 & 8.50E-03/92339 & 1.29E-02/139987 \\
\textbf{h=12} & 7.43E-03/80758 & 2.39E-04/2593 & 5.38E-05/584 & 1.29E-02/139620 & 1.29E-02/139981 & 1.44E-02/156081 \\
\textbf{h=6} & 5.58E-04/6059 & 5.87E-05/638 & 5.74E-05/624 & 1.63E-02/177183 & 1.63E-02/177219 & 1.63E-02/177219 \\
\textbf{h=3} & 3.21E-05/349 & 1.69E-05/184 & 1.69E-05/184 & 2.29E-02/248518 & 2.29E-02/248518 & 2.29E-02/248518 \\
\textbf{700k tx} & 1.63E-02/177428 & 4.61E-04/5012 & 1.66E-04/1801 & 6.43E-03/69810 & 6.44E-03/69991 & 6.44E-03/70041 \\
\textbf{d=16} & 1.54E-03/16696 & 4.72E-05/513 & 2.48E-05/269 & 2.57E-03/27924 & 2.58E-03/27996 & 2.87E-03/31216 \\
\textbf{d=32} & 1.66E-04/1806 & 2.63E-05/286 & 1.07E-05/116 & 6.40E-03/69508 & 6.41E-03/69685 & 6.40E-03/69563 \\
\textbf{d=64} & 1.69E-05/184 & 7.36E-06/80 & 6.63E-06/72 & 7.36E-03/80010 & 7.35E-03/79896 & 7.34E-03/79742 \\
\textbf{d=128} & 3.41E-06/37 & 2.02E-06/22 & 2.21E-06/24 & 8.15E-03/88542 & 8.16E-03/88668 & 8.17E-03/88723 \\
\textbf{d=256} & 1.01E-06/11 & 9.20E-07/10 & 5.52E-07/6 & 8.88E-03/96459 & 8.88E-03/96502 & 8.89E-03/96558 \\
\bottomrule
\end{tabular}
\caption{\textit{Neonpool-ETH}: Performance for n=700,000}
\label{tab:neonpooleth}
\end{table*}

\begin{figure}[ht]
\centering
\includegraphics[height=4cm, width=8.5cm]{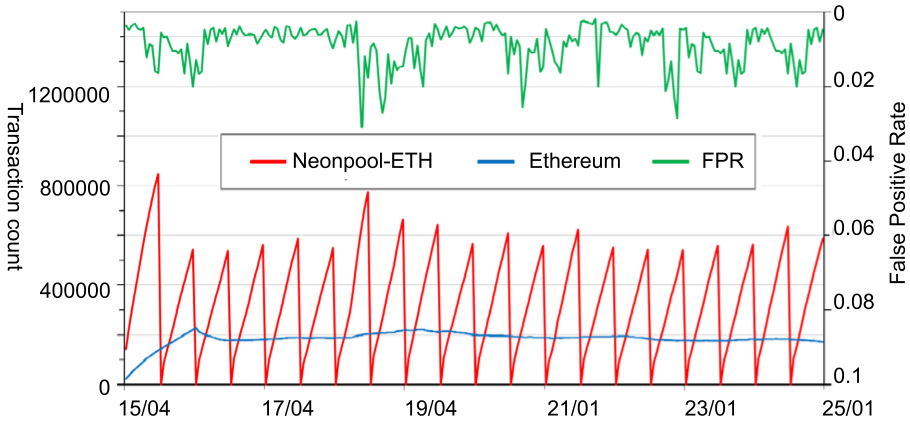}
\caption{\textit{Neonpool-ETH}: 1\,MB, 12 hours expiry}
\label{fig:neonpool_eth1}
\end{figure}

We replay transaction events in the Ethereum dataset. First, we run \textit{Neonpoool-ETH} without any transaction expiry mechanism. Thus transactions accumulate and quickly surpass the filter design capacity. Due to overloading in the filter, we get poor results. As shown in Tab.~\ref{tab:neonpooleth}, the 500\,KB, 1\,MB and 2\,MB filters report a FPR of $7.30E01$, $6.70E01$ and $6.07E01$, erroneously \textit{rejecting} around 73.0\%, 67.0\% and 60.07\% of transactions respectively. However, as no transactions are expired, the FNR is zero. 

We introduce expiry mechanisms to prevent the filters from overloading. We follow two approaches: 1. reset the bloom filter at fixed hourly intervals or once the count of transactions surpasses the number of transactions the filter was originally dimensioned for i.e. 700k in our case; 2. employ a decaying bloom filter that decrements a certain number of indices at random upon every insertion, hence mimicking expiry.

Fig.~\ref{fig:neonpool_eth1} depicts the number of transactions in \textit{Neonpool-ETH} with 1\,MB filters and 24-hour expiry along with the FPR for almost 10 days (10 million unique transactions). We also plot the corresponding number of transactions in the \textit{Ethereum txpool}, the ground truth in our evaluation. The number of transactions closely tracks the pattern in the \textit{Ethereum txpool}, with an increasing offset, as the filter retains egress transactions until the expiry interval lapses.

For instance, when the filter is cleared every 12 hours, the average FPR, at $7.43E-03$, is highest for the 500\,kB filter, reducing to $2.39E-04$ and $5.38E-05$ as the filter size increases to 1\,MB and further to 2\,MB. For the 500\,kB, 1\,MB, and 2\,MB filters, this translates to 80758 or $0.74\%$, 2593 or $0.02\%$ and 584 or $0.0569\%$ of transactions being erroneously \emph{rejected}, respectively. For each filter, there are over $1\%$ \emph{redundant} transactions due to false negatives. Tab.~\ref{tab:neonpooleth} shows that, as expected, as the expiry interval is reduced, the FPR improves while the FNR deteriorates. 

Tab.~\ref{tab:neonpooleth} also shows that when the filter is cleared every 700k transactions, the average FPR at $1.63E-02$ is highest for the 500\,kB filter and reducing to $4.61E-04$ and $1.66E-04$ as the filter size increases to 1\,MB and further to 2\,MB. For the 500\,kB, 1\,MB, and 2\,MB filters, this translates to $1.63\%$, $0.05\%$ and $0.017\%$ of transactions being erroneously \emph{rejected} due to false positives, respectively. For each filter, there are around $0.64\%$ \emph{redundant} transactions. 

Empirical false positive rates are significantly higher than the theoretical value, almost by an order of magnitude e.g. $2.39E-04$ vs $4.1E-03$ for the 1\,MB filter. This observation is consistent with \textit{Neonpool-BTC} as discussed above.

If we use a decaying bloom filter, we achieve vast improvements in terms of false positive rates. The average FPR at $3.41E-06$ is highest for the 500\,kB filter and reduces to $2.02E-06$ and $2.21E-06$ as the filter size increases to 1\,MB and further to 2\,MB. For a decay factor of 128, the 500\,kB, 1\,MB, and 2\,MB filters observe 37 or $0.0003\%$, 22 or $0.0002\%$ and 24 or $0.0002\%$ of transactions being erroneously \emph{rejected} due to false positives, respectively. For each filter, there are over $0.82\%$ \emph{redundant} transactions due to false negatives. Tab.~\ref{tab:neonpooleth} shows that by increasing the decay factor the FPR and hence the number of erroneously rejected transactions reduce. This is because the decay average meets the insertion average, and the filter reaches a stable state. However, on the flip side, the false negative rate increases. 

The \texttt{dstxFilter} which prevents double spends, can have implications denoted as $\mathbf{TP_{account}}$, $\mathbf{TN_{account}}$, $\mathbf{FP_{account}}$, and $\mathbf{FN_{account}}$. Similar to \texttt{bloomtxFilter}, a $\mathbf{TP_{account}}$ transaction should be discarded, while a $\mathbf{TN_{account}}$ transaction should be accepted. The error $\mathbf{FP_{account}}$ will lead to a genuine transaction being discarded, while $\mathbf{FN_{account}}$ will lead to accepting a transaction, the \texttt{<address,nonce>} of which has already been processed. However, circulating such transactions does not imply a double-spend, as \textit{Neonpool-ETH} and other network nodes maintain the State Trie and screen transactions in incoming blocks to prevent double-spending.

\begin{figure}
    \centering
    \includegraphics[width=8.5cm, height=4cm]{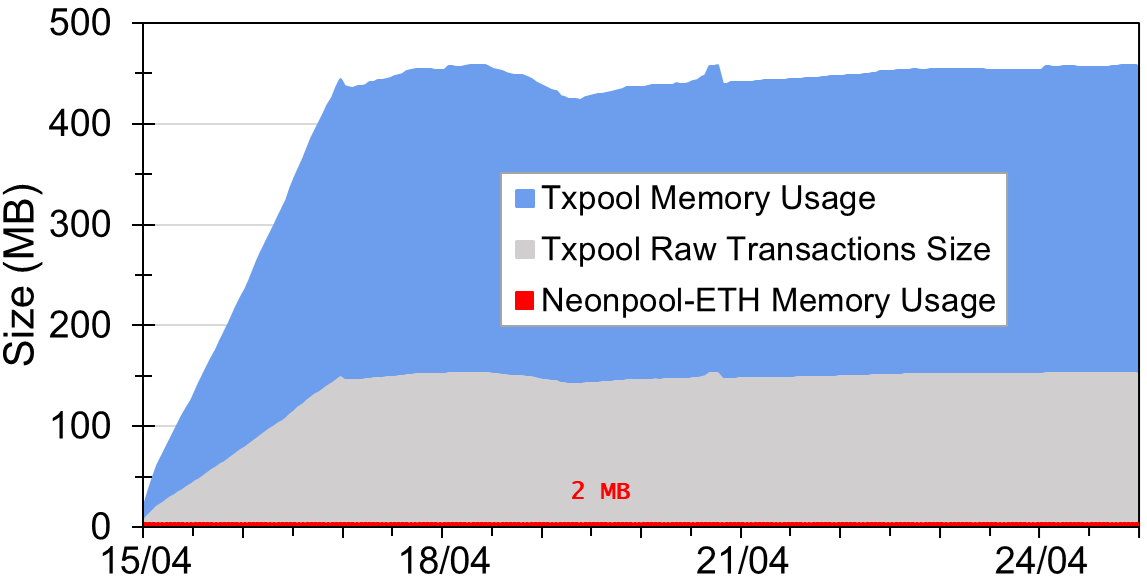}
    \caption{Memory usage of txpool and \textit{Neonpool-ETH}}
    \label{fig:neonpool_neonpoolethmemusage}
\end{figure}

Assuming each transaction is from a unique account, \texttt{dstxFilter} will have the exact dimensions as \texttt{bloomtxFilter} and consequently similar FPR. Thus, for a 1 MB \texttt{bloomtxFilter}, rejecting around 0.0005\% of valid transactions, the corresponding \texttt{dstxFilter} will also reject around 0.0005\% of valid transactions. \textbf{Neonpool-ETH achieves 99.999\% fidelity, handling 400 MB of transactions in just 2 MB, as shown in Fig.~\ref{fig:neonpool_neonpoolethmemusage}}.

\subsection{Computation time}
We conduct experiments to estimate the computation overhead of \textit{Neonpool}. The \texttt{map}-based \textit{transaction pool} in Bitcoin and Ethereum performs query, insertion, and deletion operations in $O(\log n)$ time, where $n$ is the number of stored transactions. In \textit{Neonpool}, bloom filters operate in constant time, $O(k)$, where $k$ is the number of hash functions.  

We perform simulations on \textit{Raspberry Pi 4 - Broadcom BCM2711, Quad-core Cortex-A72 64-bit @ 1.8GHz with 8 GB RAM}, and \textit{Jetson Nano Quad-core ARM Cortex-A57 64-bit @1.43 GHz MPCore processor 4 GB RAM}. Tab.~\ref{tab:neonpool_comptime} shows the computation time in microseconds ($\mu$$s$), averaged over 1E06 iterations, for querying and inserting transactions. 

\textit{Neonpool} is designed to avoid increasing computational demands, which directly translates to energy consumption. In fact, \textit{Neonpool} reduces computation time compared to traditional transaction pools:

In Bitcoin, processing an incoming transaction begins with querying the transaction's hash, followed by querying its inputs\footnote{For simplicity, we assume each transaction has a single input. As the number of inputs increases, the number of queries increases proportionally in both Bitcoin Core and Neonpool-BTC.}. If the transaction is new and has valid inputs, it is then added to the map along with its inputs. This process involves two queries and two insert operations. Similarly, in Neonpool-BTC, the transaction’s hash is queried first, followed by a query for its inputs. If the transaction is new and has valid inputs, it is added to Neonpool-BTC along with its inputs. This also involves two queries and two insert operations. The total computational cost for this process can be expressed as: $2\times(query\: time + insert\:time)$.

Bitcoin, for querying and inserting a single transaction and its inputs, takes 20.2\,$\mu$$s$ ($2\times (1.8+8.3)$ $\mu$$s$) on a \textit{Raspberry Pi 4} and 27.2\,$\mu$$s$ on a \textit{Jetson Nano} on average. \textit{Neonpool-BTC} with \texttt{bloomtxFilter} and \texttt{dstxFilter} dimensioned at 1 MB with k=14 hash functions, cumulatively for query and insert, takes 20.8\,$\mu$$s$ ($2\times (4.4+6.0)$ $\mu$$s$) on a \textit{Raspberry Pi 4} and 26\,$\mu$$s$ on a \textit{Jetson Nano}, on average.

Similarly, in Ethereum, processing starts with querying the transaction’s hash, followed by querying the nonce if the transaction is fresh. If the transaction is new and has a valid nonce, it is added to the txpool, and the nonce is updated. This process also requires two queries and two insert operations. The computational cost for this process is the same: $2\times(query\:time + insert\: time)$.

Ethereum, for querying and inserting a single transaction and the state information, takes 17.4\,$\mu$$s$ ($2\times (6.9+1.8)$ $\mu$$s$) on a \textit{Raspberry Pi 4} and 20.6\,$\mu$$s$ on a \textit{Jetson Nano} on average. Similarly, \textit{Neonpool-ETH} with \texttt{bloomtxFilter} and \texttt{dstxFilter}, dimensioned at 1 MB and k=8 hash functions, cumulatively for query and insert, will take 12.4\,$\mu$$s$ ($2\times (2.9+3.3)$ $\mu$$s$) on a \textit{Raspberry Pi 4} and 14.4\,$\mu$$s$ on a \textit{Jetson Nano}, on average.

Thus for practical values of $k$ \textit{Neonpool} does not increase computation load. \textit{Neonpool} can scale to support cryptocurrencies throughout to the order of thousands of transactions per second (tps). However, the current throughput for Bitcoin and Ethereum is around 3-7 tps and 15-20 tps, respectively.

\begin{table}[htbp]
\setlength{\tabcolsep}{2pt}
\begin{tabular}{lrrrrrrrr}
\toprule
\multicolumn{4}{c}{\textbf{Raspberry Pi 4}} & \multicolumn{4}{c}{\textbf{Jetson Nano}} \\
\midrule
& \multicolumn{1}{c}{Bitcoin} & \multicolumn{3}{c}{\textit{Neonpool-BTC}} & \multicolumn{1}{c}{Bitcoin} & \multicolumn{3}{c}{\textit{Neonpool-BTC}} \\
& & k=7 & k=14 & k=28 &   & k=7 & k=14 & k=28 \\
Query & 1.8 & 2.6 & 4.4 & 9.6 & 2.4 & 3.5 & 6.0 & 12.4 \\
Insert & 8.3 & 2.9 & 6.0 & 12.6 & 11.2 & 4.1 & 7.0 & 15.4 \\
\midrule
& \multicolumn{1}{c}{Ethereum} & \multicolumn{3}{c}{\textit{Neonpool-ETH}} & \multicolumn{1}{c}{Ethereum} & \multicolumn{3}{c}{\textit{Neonpool-ETH}} \\
&  & k=4 & k=8 & k=16 &  & k=4 & k=8 & k=16 \\
Query & 1.8  & 1.6  & 2.9  & 4.3  & 2.1  & 1.9  & 3.4  & 5.1 \\
Insert & 6.9  & 1.8  & 3.3  & 5.1  & 8.2  & 2.1  & 3.8  & 5.8 \\
\midrule
\end{tabular}
\caption{Query/Insert time ($\mu$s)}
 \label{tab:neonpool_comptime}
\end{table}

\subsection{Security analysis}
Here, we establish the security of \textit{Neonpool}, focusing on two main aspects: 1. whether errors made by \textit{Neonpool} compromise its security or that of the broader network; and 2. \textit{Neonpool}'s resilience to adversarial attacks.

The network-level impact of false positives vanishes at the network level, i.e., effectively negligible. Each node initializes its Bloom filters with a unique, random 128-bit salt, ensuring independence between filters across nodes. Consequently, false positives at one node are statistically independent of those at other nodes. For instance, with a Bloom filter accuracy of 99.99\% (corresponding to a false positive rate of 0.0001 per \textit{Neonpool} node), the probability of two nodes erroneously dropping the same transaction is $(0.0001)^2$, an exceedingly low likelihood.

Furthermore, research indicates that while transaction pools across nodes are not entirely identical, they exhibit a remarkable 99\% similarity in their contents~\cite{dae2020examining}. This high consistency underscores that, despite occasional discrepancies introduced by Bloom filter false positives or false negatives, the overall integrity of transaction pool contents across the network remains robust, ensuring reliable transaction propagation.

In Bitcoin and Ethereum, there is an established practice of limiting transaction pool sizes and managing overflow by rejecting or expiring excess transactions. For instance, Bitcoin’s default mempool size is capped at 300 MB, while Ethereum employs a default limit of 4096 transactions in the transaction pool, with surplus transactions being evicted~\cite{mempoollimit}~\cite{srccodetxpoolgo}. Users can customize transaction pool policies or disable transaction pools entirely to accommodate low-memory or low-computation environments.

Additionally, an adversary may: 1. trigger false positives to censor specific transactions; 2. craft invalid transactions that evade verification and validation; 3. generate spam. 

Literature shows that any bloom filter can be efficiently transformed to be adversarial resilient by applying a pseudo-random permutation of the input \cite{naor2019bloom} i.e. applying a sufficiently large (128-bit) random salt before forwarding it to the bloom filter. This change requires little overhead and randomizes the adversary’s queries by applying a pseudorandom permutation to them; then, we may consider the transactions sent by the attacker as random and not as chosen adaptively by the adversary. It is also recommended that a node regenerate its 128-bit random salt every time a new \texttt{bloomtxFilter} or \texttt{dstxFilter} is generated. Thus, the adversary only has oracle access to the bloom filter and does not know its contents or seed.

We situate our assumption within established practices in the cryptocurrency ecosystem and light client security models:
Secure random seed generation is essential for the security of Bitcoin, Ethereum, and the broader crypto ecosystem. It underpins wallet key generation, ensuring unique private keys. Random seeds also facilitate contract deployment through ECDSA nonce generation, enable ephemeral session keys for secure communication, and support multi-signature wallets by creating unique key shares for participants.

As identified by Chatzigiannis et al.~\cite{chatzigiannis2022sok}, there are several common assumptions that underpin light client designs, including trusted genesis block, reliable consensus, secure underlying cryptographic primitives, weak synchrony (i.e., no long network partitions), trusted setup, peer-to-peer communication for relaying information, and rational behaviour of participants. Secure underlying cryptographic primitives and trusted setups are of particular interest to us.

In this context, our assumption about Oracle access to the Bloom filter aligns with these principles. Secure generation and protection of the 128-bit random seed are essential for the Bloom filter's adversarial resilience and are consistent with best practices in decentralized systems. The cryptographic strength and manipulation resistance of the hash functions in Bloom filters fundamentally depend on this secure initialization and randomization.

Secondly, Eve may craft invalid transactions to evade verification and validation, i.e., attempt double-spending. \textit{Neonpool} preserves the verification and validation mechanisms of Bitcoin and Ethereum. Regarding transactions with conflicting inputs or out-of-order nonce, typically, nodes accept and forward the first seen transaction, and the first seen can differ for nodes. It is the job of miners not to add conflicting transactions to a block and the network nodes to screen incoming blocks. Since \textit{Neonpool} maintains complete UTXO and Trie information, \textit{Neonpool} nodes will reject blocks that include double-spend transactions.

Thirdly, Eve might launch a dust or spam attack or replay transactions. A 2015 Bitcoin spam campaign swelled the transaction pool to nearly 1 GB, crashing 10\% of nodes, mostly memory-constrained like Raspberry Pi. \textit{Neonpool} can withstand such attacks by recursively generating additional bloom filters on demand, as described in section~\ref{sec:neonpool_proposed}.

Replay transactions are rejected. Already seen transactions will trigger a positive in \textit{Neonpool}, indicating that the transaction is already present and thus will be dropped.

\subsection{Summary}
The unconfirmed transaction pool plays a critical role in verifying, storing, and disseminating transactions while they await inclusion in a block. We present \textit{Neonpool}, a novel transaction pool construction for cryptocurrencies that stores transaction fingerprints via bloom filters instead of storing complete transactions via map data structures. We perform benchmarks using unique Bitcoin and Ethereum datasets comprising approximately 10 million unique transactions. We achieve up to two orders of magnitude reduction in memory consumption, fingerprinting up to 400 megabytes of data in as low as 2 MB while maintaining a verification and forwarding accuracy exceeding 99.99\%, with a slight increase in computation load. We also demonstrate its adversarial resilience. We summarize our findings in Tab.~\ref{tab:neonpool_summary}.

Due to their function in the network, Neonpool node operators are not required to store full transactions. This conceptually resembles Bitcoin Core’s built-in "pruned node" option, which reduces hard disk requirements by storing only a few recent blocks instead of the full blockchain on disk while still contributing to the network’s footprint and health by validating and forwarding transactions.

Such full-node users operate nodes primarily to contribute to the Bitcoin network out of a sense of community or altruism, similar to how people operate nodes for the Tor network. Neonpool lowers the barrier to entry for such non-mining full nodes, offering them greater control over the memory resources they allocate for the transaction pool.

Neonpool provides full-node operators with significant flexibility to optimize resource usage by adjusting memory and computation allocations to balance efficiency and functionality. Lightweight devices, like IoT nodes, can store only transaction fingerprints in Bloom filters, while full nodes participating in mining can still retain a subset of full transactions to propose a block.

\begin{table}[htbp]
\begin{tabular}{lcc}
Transaction(s) & Bitcoin / Ethereum & \textit{Neonpool}- BTC/ETH \\
\midrule
Storage & complete & fingerprint \\
Data Structure & map-based & bloomtxFilter, \\
  & mempool/txpool & dstxFilter \\
Memory Usage & up to 400 MB & 2\,MB \\
Verification  & Yes & Yes \\
Inventory & Yes & Probabilistic (99.99\%) \\
\& Propagation & & \\
\midrule
\end{tabular}
\caption{\textit{Neonpool} vs Bitcoin/Ethereum}
    \label{tab:neonpool_summary}
\end{table}

\section{Prior work}
\label{sec:neonpool_prior}
Our work relates to two main bodies of research: lightweight clients and transaction pool management.

\subsection{Lightweight clients}
Our work relates to two main bodies of research: lightweight clients and transaction pool management.
Our work has a tangential relationship with existing light clients. While current light clients effectively address specific resource constraints, such as storage, computation, or bandwidth, none focus on reducing memory consumption in the transaction pool—a critical yet often overlooked challenge. Consequently, direct comparisons with these approaches are not feasible. However, this distinction helps position our work within the broader spectrum of light clients. Notably, these light-client solutions are orthogonal to our approach and can be deployed alongside Neonpool if needed. We present key contributions from the literature on lightweight clients, emphasizing that no existing solutions propose a lightweight version of the transaction pool.

\begin{table}[b]
\setlength{\tabcolsep}{2pt}
\centering
\begin{tabular}{llllll}
Scheme & Target & Consensus & Model & Integration & Primitive(s) \\
\midrule
SPV\cite{nakamoto2008} & blocks & Any & Any & Yes & - \\
NiPoPoW\cite{kiayias2020non} & blocks & PoW & UTXO & Mod & NiPoPoWs \\
Flyclient\cite{bunz2020flyclient} & blocks & PoW & UTXO & Mod & MMR \\
PoNW\cite{kattis2020proof} & blocks & PoW & UTXO & New/Mod & SNARKs \\
EdraX\cite{chepurnoy2018edrax} & state & Any & Any & New/Mod & SparseMT, \\
                               &        &     &     &         &  Dist.VC\\
Ethanos\cite{kim2021ethanos} & state & PoS & Account & Mod & - \\
\textit{Neonpool} & txpool & Any & Any & Mod & Bloom filters \\
\midrule
  \end{tabular}
    \caption{\textit{Neonpool} in the light client spectrum}
    \label{tab:neonpool_comparison}
\end{table}

\textbf{Reducing blockchain overheads:} Satoshi Nakamoto introduced Simplified Payment Verification (SPV) clients as a lightweight client, which requires download of only block headers and select blocks to verify transactions~\cite{nakamoto2008}. However, these scales linearly: Ethereum's SPV client storage exceeds 10 GB as of July 2023~\cite{ethclients}.

Pruned nodes retain only a recent subset of the blockchain. While they offer robust security, they cannot bootstrap new nodes. Ultra-light clients of this type depend on trusted full nodes since they cannot verify transactions independently, leading to security and privacy concerns.

\textbf{Reducing bootstrapping costs:} Kiayias et al.~\cite{kiayias2020non} introduced sublinear storage complexity in SPV clients via skip lists, termed noninteractive proofs of proof-of-work (NIPoPoW). This solution checks for high-difficulty previous blocks. Verifying a logarithmic number of these suffices to ensure security for the whole chain. However, this solution is only practical in an honest network with fixed difficulty, unlike most cryptocurrencies with variable block difficulty.

FlyClient~\cite{bunz2020flyclient} achieves logarithmic complexity, using Merkle Mountain Range Commitments for memory improvements and a random block sampling protocol to ensure security. This solution works even if parts of the network are adversarial and have variable block difficulty. However, NiPoPoW and FlyClient still require linear resources, and verifying transactions remains costly, as each verified transaction also requires downloading the corresponding block.  

TXCHAIN~\cite{zamyatin2020txchain} addresses this issue using contingent transaction aggregation to compress transaction inclusion proofs. Proof of Necessary Work~\cite{kattis2020proof} performs necessary system verification within the proof-of-work computation, utilizing SNARKs and Pederson hash.

\textbf{State optimizations:} Bitcoin's UTXO and Ethereum's state trie occupy tens of gigabytes, prompting proposals for more efficient representations: Utreexo~\cite{dryja2019utreexo} and BZIP~\cite{jiang2019bzip} recommend representing the UTXO using hash-based accumulators and lossless compression methods. Dietcoin~\cite{frey2019dietcoin} splits UTXO into shards, while EDRAX~\cite{chepurnoy2018edrax} uses sparse Merkle trees for UTXO and vector commitments for the state trie. Ethanos downsizes the state trie by periodically emptying idle accounts~\cite{kim2021ethanos}.

\textbf{Network optimizations:} Graphene uses bloom filters to reduce network bandwidth in block reconciliation~\cite{ozisik2019graphene}. Anas et al. recommend increasing the orphan pool size from 100 to 1000, reducing their overhead by 17\%. Other works propose lightweight transaction broadcasting: Strokkur uses rateless erasure LT codes~\cite{kattis2020proof}, Erlay combines limited flooding with intermittent reconciliation~\cite{naumenko2019erlay}, and Shrec employs an efficient low-collision hybrid hashing scheme~\cite{han2020shrec}.

\subsection{Transaction pool}
This remains a neglected area in the research literature, with earlier works focusing on mitigating spam, dust, and DDoS attacks by filtering malicious transactions.

Baqer et al. were the first to emphasize the transaction pool's significance in their analysis of a 2015 Bitcoin stress test~\cite{baqer2016stressing}. This attack expanded the pool to nearly 1\,GB, reportedly causing 10\% of Bitcoin nodes to crash. They classified 23\% of transactions as spam using clustering techniques and proposed spam filtering. However, they warned about the risks of misclassifying legitimate transactions: even a 1-2\% false positive rate could create a self-inflicted DoS attack. Additionally, attackers could manipulate transaction attributes to bypass filters if mechanisms were exposed.

Subsequent works focused on transaction filtering. Saad et al. introduced Contra, which filters spam based on age and fee thresholds~\cite{saad2020contra}. Configuring these thresholds presents tradeoffs: high thresholds risk false positives, while low thresholds risk false negatives. Their modelling estimates 60\% accuracy, recommending dynamically increasing block size to accommodate dust transactions and deter spammers, though results lack real-world validation. Wang et al. proposed an \textit{Anti-dust} solution, analyzing Bitcoin transactions (2009-2017) with a Gaussian model~\cite{wang2018anti}. Transactions below a threshold were classified as spam and placed in a dust pool, later moved to the mempool if space allowed. Overflowing transactions were discarded. Simulations showed dust transactions increased validation time from 200 to 25,000 seconds, while Anti-dust reduced it to 215 seconds.

Eduardo et al. simulated dust attacks on Ethereum using 2 million genuine transactions and synthetic ones~\cite{eduardo2021fighting}. They found dust attacks extended transaction pending time by over 42\%. Using machine learning, they achieved 94\% accuracy in identifying under-priced potential DoS attack transactions. DETER highlighted Ethereum-specific vulnerabilities, describing DETER-X and DETER-Z attacks that evict legitimate transactions, delay processing, and reduce miner revenue~\cite{li2021deter}. Proposed heuristics mitigate these attacks by regulating txpool entry and eviction.

To the best of our knowledge, Neonpool is the first optimization technique that re-architects the transaction pool of a cryptocurrency from an optimization perspective, specifically aiming to reduce the local memory consumption of the transaction pool. Earlier works primarily focus on mitigating spam, dust, and DDoS attacks by filtering out malicious or low-value transactions. They do this by introducing additional modules to the transaction pool while preserving the transaction pool structure itself. These approaches increase resource usage, as implementing real-time filtering (via statistical techniques or machine learning) or maintaining separate pools for spam imposes significant computational and memory costs, which are not adequately investigated.

Our work does not focus on spam filtering but prioritizes reducing memory usage and enhancing transaction pool resilience for larger traffic flows. As a result, a direct comparison with these techniques is not possible. Filtering solutions are orthogonal to our overall approach and can be deployed against \textit{Neonpool} if required.

Our work is directly compared with the reference implementations of Bitcoin Core and Ethereum, as detailed in \S\ref{sec:neonpool_results}.

\section{Conclusions and Future Work}
\label{sec:neonpool_conclusion}
Our work introduces a promising new direction in the domain of light clients, scalability solutions, and improving the health of cryptocurrency networks. \textit{Neonpool} proposes a novel transaction pool design based on bloom filter variants and achieves a remarkable reduction of up to 200x in memory usage while maintaining a verification and forwarding accuracy of over 99.99\%. This breakthrough makes it a viable solution for supporting resource-constrained devices, such as browsers, smartphones, systems-on-a-chip, mobile, and IoT devices, to perform full-node functions effectively. Additionally, \textit{Neonpool} does not require a hard fork.

Our results highlight \textit{Neonpool}’s potential and provide a foundation for further exploration in several directions.

Our work pioneers investigating the suitability of probabilistic data structures for transaction pool construction. While over a dozen Bloom filter variants exist in the literature, we begin with the simplest ones that meet our requirements and are widely understood. Future research will evaluate alternative probabilistic data structures, such as counting Bloom filters and cuckoo filters, to explore improved trade-offs in error rates, memory and computational overhead.

Another focus of future work is to assess \textit{Neonpool}’s scalability under varying network conditions, transaction loads, and scenarios such as spam and dust attacks. Future experiments will aim to log and analyze attack patterns, examine the relationship between transaction fee spikes and network congestion, and design robust defences.

Finally, we plan to introduce \textit{Neonpool}’s design to the cryptocurrency community by initiating a Bitcoin Improvement Proposal (BIP) and an Ethereum Improvement Proposal (EIP), to facilitate \textit{Neonpool}’s live deployment and integration into existing blockchain ecosystems.

Neonpool’s approach can be extended to other cryptocurrencies with minimal modifications. For instance, \textit{Neonpool-BTC} can be adapted for UTXO-based systems, and \textit{Neonpool-ETH} for account-based cryptocurrencies by adjusting parameters like transaction expiry time and filter size based on network conditions and block time. These adaptations enable generic client-side optimizations, making \textit{Neonpool} broadly applicable to lightweight devices across different distributed ledger technologies (DLTs). We are currently preparing datasets for alternative currencies, including Litecoin and Solana, to evaluate \textit{Neonpool}'s adaptability and effectiveness across diverse blockchain architectures.

\end{document}